\documentclass{elsarticle}
\journal{Astroparticle Physics}
\begin{document}
\begin{frontmatter}
\title{On the origin of galactic cosmic rays}

\author[Fian]{Ya. N. Istomin},
\ead{istomin@lpi.ru}

\address[Fian]{P.~N.~Lebedev Physical Institute,
Leninsky Prospect 53, Moscow, 119991 Russia}

\begin{abstract}
It is shown that the relativistic jet, emitted from the center of
the Galaxy during its activity, possessed power and
energy spectrum of accelerated protons sufficient to explain the
current cosmic rays distribution in the Galaxy. Proton
acceleration takes place on the light cylinder surface formed by
the rotation of a massive black hole carring into rotation the
radial magnetic field and the magnetosphere. Observed in gamma,
x-ray and radio bands bubbles above and below the galactic plane
can be remnants of this bipolar get. The size of the bubble
defines the time of the jet's start, $\simeq 2.4\cdot 10^7$ years
ago. The jet worked more than $10^7$ years, but less than
$2.4\cdot10^7$ years.

\end{abstract}


\end{frontmatter}

\section{Introduction}
The traditional point of view on the origin of cosmic rays in the
Galaxy is the concept of acceleration of charged particles at
fronts of shocks from supernova explosions. Arguments in favour of
this mechanism are sufficient mechanical energy that is released
when the supernova explodes, as well as universal index of power
law spectrum of particles, accelerated by strong shocks. Total
power of cosmic ray sources in order to maintain their observed
density of energy is $5\cdot 10^{40} erg/s$, which equals
approximately 15\% of the kinetic energy of supernova explosions.
When the gas compression in a shock is equal to 4, the index of
the power law energy spectrum of accelerated particles is equal to -2, $N
(E)\propto E^{-2}$. It is in a good agreement with observed cosmic
ray spectrum at energies $E<3\cdot 10^{15} eV$. Beginning from the
first works by Krymskii, 1977; Bell, 1978; Blandford \& Ostriker,
1978, who proposed the mechanism of acceleration of charged
particles on fronts of shocks propagating in the turbulent
environment, much progress has been made to explain the observed
characteristics of galactic cosmic rays in the belief that they
are accelerated at the front of shocks.

On the other hand, there is no objections to generate
galactic cosmic rays in one source in the Galaxy (Ptuskin \& Khazan, 1981).
This potential source can be the center of the Galaxy, which is the
massive black hole of $M\simeq 4\cdot 10^6 M_{\odot}$ mass.
And while the luminosity of Sgr A* is small now, it is only $10^{36}$ erg/s,
in the past the center could be much brighter because its Eddington
luminosity equals $L_{Edd} = 5.2\cdot 10^{44}$ erg/s. On the past
activity of the center of the Galaxy shows newly discovered by Fermi Gamma-ray
Space Telescope above and below the galactic plane big bubbles emitting
gamma radiation in the range of $0.1-1000$ GeV (Su et al., 2010).
Such formations
was previously observed in the x-ray range $(1.5-2)$ KeV by ROSAT All-Sky
Survey (Snowden et al., 1997) and in the microwave range $(20 -40)$ GHz by WMAP
(Finkbeiner, 2004).
Estimated energy stored in bubbles is of $10^{54}-10^{55}$
erg (Sofue, 2000). As we will see below, bubbles of a relativistic gas
could be formed by the jet, emitted from surroundings of the massive
black hole.
Here we provide an alternative mechanism of origin galactic cosmic rays,
in which the nucleus of the Galaxy in the active phase injected the
relativistic jet, which was the source of cosmic rays.

In the following sections we will calculate the power of the jet
and the energy spectrum of protons in the relativistic jet, as
well as describe the remnants of the relativistic jet injected
from the center of the Galaxy, having the form of bubbles above
and below the galactic plane. In the final section we will discuss
correspondence of the scenario of the galactic cosmic rays origin
provided with cosmic rays characteristics observed.

\section{Relativistic jet}
Sources of energy of active galactic nuclei are the accretion on a
massive black hole, in which the gravitational energy of a
falling gas transforms into radiation and heat, as well as the
rotation of a black hole. Mechanism of extraction of energy and
angular momentum from the black hole is called as the mechanism of
Blandford-Znajek (1977). The energy of a rotating black hole is a
large value, $E_{rot} = Mr_H^2\Omega_H^2/2 = a^2Mc^2/8 = 2.25\cdot
10^{53} a^2 (M/M_\odot)$ erg. For the Galaxy $E_{rot}\simeq 9\cdot
10^{59} a^2$ erg. Here we have introduced the dimensionless
parameter of $a$, describing rotation of the black hole, $a =
Jc/M^2G, \, a<1$. $J$ is the black hole angular momentum, $G$ is
the gravitational constant. Angular velocity of rotation of a
black hole is proportional to the value of $a$, $\Omega_H =
ac/2r_H$, $r_H$ is the gravitational radius of not rotating black
hole $(a = 0)$, $r_H =2MG/c^2$. Energy extraction is possible when
there is a poloidal magnetic field $B$ near the black hole
horizon. In this case, rotating black hole acts as a Dynamo
machine, creating a voltage $U = f_H\Omega_H/2\pi c$ (Landau \&
Lifshits, 1984). The value of $f_H$ is the flux of the poloidal
magnetic field reaching the horizon of a black hole, $f_H\simeq
\pi Br_H^2$. Voltage $U$ generates the electric current $I = U/(R
+ R_H)$, which on the one hand is closed on the horizon of a
black hole that has the resistance  $R_H = 4\pi/c\simeq 377$ ohm.
Resistance of the outer part of the current loop is $R$. Thus, the
power, extracting from a rotating black hole, is $L = RI^2 =
U^2R/(R + R_H)^2 = a^2B^2r_H^2R/16 (R + R_H)^2$, and reaches the
maximum $L_m$ at $R = R_H$, $L_m = a^2B^2r_H^2c/256\pi$. The value
of $L$ is proportional to the energy of the poloidal magnetic
field near a black hole and can reach the Eddington luminosity at
sufficiently large magnetic fields $B\simeq 10^6 a^{-1}$ Gauss in
the center of the Galaxy. This field is accumulated near the
horizon of a black hole in the process of accretion of a
disk matter in which the magnetic field is frozen. Thus, for the
effective work of the mechanism of Blandford-Znajek an accretion
disk around a massive black hole is required, not as a source of
the energy, but as the agent bearing the magnetic field to a
black hole. In addition, the electric current $I$ flows in the
disk, this is the part of the current loop: in the disk, in the black
hole horizon, then in the jet, closing at large distances in the
interstellar matter (see Figure 1).
\begin{figure}
\begin{center}
\includegraphics[width=8cm]{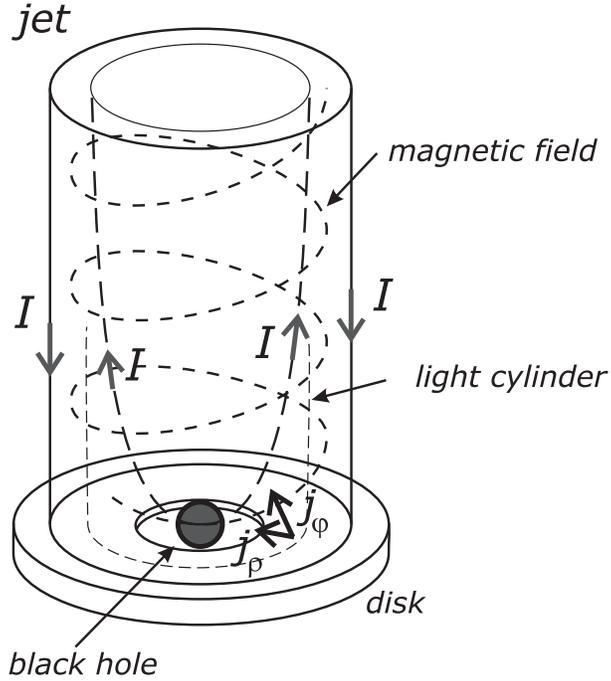}
\end{center}
\caption {Configuration of the magnetic field and electric
currents in the jet and in the disk.}
\end{figure}

In the disk, in addition to the radial electric current
$I_\rho = \int j_\rho ds =-I$, stronger toroidal
current $j_\phi$, $j_\phi\simeq 10^2
j_\rho$, flows also (Istomin \& Sol, 2011),
it generates the radial magnetic field $B$.

The rotating black hole brings into rotation the radial magnetic
field in the magnetosphere of a black hole above and below the
disk. Angular velocity of rotation of the magnetic field lines,
the same as rotation of the magnetospheric plasma, $\Omega_F$, is
proportional to the angular velocity of rotation of the black hole
$\Omega_H$, $\Omega_F = \Omega_H R/(R + R_H)$ (Thorne et al.,
1986). Plasma rotation is the drift motion in crossed radial
magnetic field and electric field of plasma polarization. Thus,
there appears so-called the light cylinder surface in the black
hole magnetosphere, where the magnitude of the electric field is
compared with that of the magnetic field and the rotation velocity
approaches the speed of light $c$. The radius of the light surface
is  $r_L=c/\Omega_F=2a^{-1}r_H(R+R_H)/R>r_H$. On the light surface
charged particles get considerable energy and angular momentum of
rotation. Energy density of particles on the light surface is
compared with the energy density of the electromagnetic field
$(E_L^2+B_L^2)/8\pi=B_L^2/4\pi$ (Istomin, 2010). All energy passes
to protons, $\gamma = {\cal E}_p/m_pc^2 > >1$. Electrons are
practically not accelerated due to
large synchrotron losses in a strong magnetic field (Istomin \& Sol, 2009). 
Energetic
protons, accelerated near the light surface, and whose energy is
mainly in the azimuthal motion, create the jet. Jet's power is
$L_J = B^2r_H^2 c (\omega_ {cH}r_H/c)^{-1/4}/2$ (Istomin \& Sol,
2011). Here $\omega_{cH}$ is the non relativistic cyclotron
frequency of protons in the magnetic field near the black hole,
$\omega_{cH} = eB/m_pc$. Jet arises when the power extracted from
the rotating black hole $L$ becomes greater than the jet power
$L_J$,  $L>L_J$. This imposes a limitation on the value of the
magnetic field
\begin{equation}
\frac{\omega_{cH}r_H}{c}\ge(128\pi)^4 a^{-8}\left[\frac{(R+R_H)^2}{4RR_H}
\right]^4.
\end{equation}
For $R=R_H$ it gives
\begin{equation}
B\ge 2.7\cdot 10^{11}a^{-8}\frac{M_\odot}{M}{\rm Gauss}.
\end{equation}
For the center of the Galaxy, the magnetic field must satisfy the
condition $B\ge 6.75\cdot 10^4 a^{-8}$ Gauss. We see that to
generate a jet, less massive black holes should have a stronger
poloidal magnetic field near the horizon, $B\propto M^{ -1}$. In
addition, rotation must be fast, close to the critical value
of $a\simeq 1$, because of strong dependence of the
expression (2) on $a$. It should also be noted that the resistance of the
external current loop $R$, on which the jet power $L_J$ depends, is
not the ohmic one $R_c$, which turns out to be small, $R_c < <
R_H$ (Istomin \& Sol, 2011), but is the effective resistance
$R_J$, which can be attributed to the jet, receiving energy from
the rotating black hole. If $L = L_J$ the resistance of the jet is
$R_J = R_H$. Under the equality in the expression (1) when a
rotating black hole can begin to generate a jet, the expression
for the jet's power becomes universal
\begin{equation}
L_J=2^{48}\pi^7m_pc^2\left(\frac{m_pc^3}{e^2}\right)a^{-14}=2.5\cdot 10^{41}
a^{-14} \, erg/s.
\end{equation}
All jet energy are in the energy of protons, and, as we can see, is
sufficient for the production of galactic cosmic rays.

\section{Energy spectrum of fast particles in jet}

Istomin and Sol (2009) had shown that on the light surface,
produced by the rotating radial magnetic field, which is carried
into rotation by a black hole, protons gain considerable energy.
The Lorentz factor $\gamma$ becomes equal to $\gamma =
(\gamma_0\gamma_i)^{1/2}$. The value of $\gamma_i$ is the Lorentz
factor of particles in the magnetosphere of a black hole before
crossing the light surface. And the value of $\gamma_0$ is the
maximum of the Lorentz factor, which could be achieved by a
particle in this acceleration mechanism, $\gamma_0 =
\omega_{cL}/\Omega_F$. Here $\omega_{cL}$ is the cyclotron
frequency of rotation of protons in the poloidal magnetic field
near the light surface. When $\gamma = \gamma_0$ the cyclotron
radius of a proton is compared with the radius of the light
surface. For non relativistic particles of the black hole
magnetosphere, $\gamma_i\simeq 1$, the Lorentz factor of
accelerated particles is equal to $\gamma = \gamma_0^{1/2} =
(\omega_{cL}/\Omega_F)^{1/2}$. Crossing the light surface at
different distances $z$ from the accretion disk plane, particles
gain different energies, since the magnetic field decreases with
distance from the black hole. For a radial magnetic field
$B\propto (z^2 + r_L^2)^{-1}$. Thus, $\gamma =
\gamma_m(1+z^2/r_L^2)^{-1/2}$, where $\gamma_m$ is the maximal
Lorentz factor of accelerated particles near the accretion disk.
Accelerated protons of the jet are collected from various parts of
the light cylinder surface of $r_L$ radius, but located at
different distances $z$. Therefore, the number of particles is
$dN\propto ndz$, where $n$ is the density of protons in the
magnetosphere near the light surface. Connecting values $z$ and
$\gamma$, we get
$$
\frac{dz}{r_L}=-\frac{\gamma_m d\gamma}{\gamma^2(1-\gamma^2/\gamma_m^2)^{1/2}}.
$$
Considering that the vertical size of the magnetosphere is larger than the
light surface radius, the density $n$ can be taken as constant.
As a result we get the
distribution function of relativistic protons in the jet,
$F(\gamma) = dN/d\gamma$,
\begin{equation}
F(\gamma)=const\cdot \gamma^{-2}(1-\gamma^2/\gamma_m^2)^{-1/2},
\, \gamma<\gamma_m.
\end{equation}
We see that in the range $\gamma<<\gamma_m$ the spectrum of relativistic
protons is the power law spectrum with the index
-2. This spectrum is observed in gamma radiation
from bubbles above and below the Galactic plane by Fermi Gamma-ray Space
Telescope (Su et al., 2010). Considering that the gamma radiation occurs
due to collisions of relativistic protons with interstellar gas through
meson production (Crocker \& Aharonian, 2011), and the distribution of
photons is similar to the distribution of protons, one can conclude
that the jet from the center of the Galaxy actually existed, and bubbles are
filled with relativistic protons of the jet.
The value of $\gamma_m$ is (Istomin \& Sol, 2011)
\begin{equation}
\gamma_m = \left(\frac{\omega_{cH}r_H}{c}\right)^{1/2},
\end{equation}
and for $R = R_H$ equals (see the expression (1))
$$
\gamma_m=(128\pi)^2 a^{-4}=1.6\cdot 10^5 a^{-4}.
$$
The spectrum of protons (4) breaks at $\gamma = \gamma_m$ and has
there the root singularity (integrable) that is smoothed
considering the thermal dispersion of particles in the
magnetosphere of the black hole, $\Delta\gamma_i = {\cal
E}_p/m_pc^2$. The distribution (4) is shown on Figure 2. The value
of $\gamma_m$ is chosen to be equal to the Lorentz factor of the
break in the observed spectrum of cosmic rays at the energy
$E=3\cdot 10^{15}$ GeV, $\gamma_m=3.2\cdot 10^6$. This corresponds
to the rotation parameter $a = 0.47$. The power of the jet (3) is
$L_J\simeq 8.9\cdot 10^{45}$ erg/s. That is in agreement with
estimated from observations powers of jets ejected from active
galactic nuclei--$10^{45}-10^{46}$ erg/s (Mao-Li et al., 2008).
\begin{figure}
\begin{center}
\includegraphics [width = 8 cm] {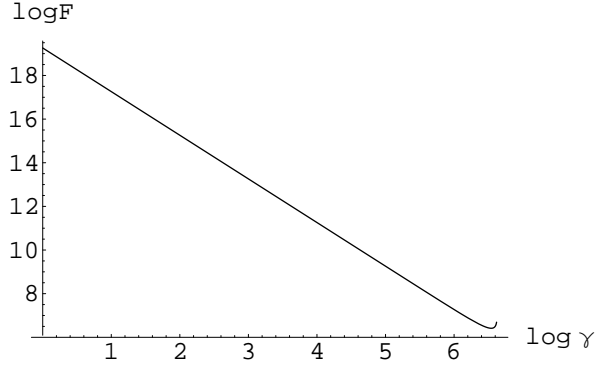}
\end{center}
\caption {Distribution function of relativistic protons
in the jet, $\gamma <\gamma_m$. The value of $\gamma_m$ corresponds to the
break in the spectrum of the cosmic ray. The slope is
equal to -2.}
\end{figure}

Relativistic protons with the spectrum (4) are formed from thermal
particles of the black hole magnetosphere, $\gamma_i\simeq 1$. But
in addition to thermal particles in the magnetosphere there can
exist accelerated protons. The turbulent motion of the accreting
disk matter in the presence of the frozen magnetic field leads to
acceleration of particles, which have the power law energy
spectrum, $f(\gamma) = const\cdot\gamma^{-\beta}, \,
\gamma<\gamma_1, \, \beta\simeq 1$ (Istomin \& Sol, 2009). The
disk must be turbulent to provide for the abnormal gas transport.
Getting onto the light surface, accelerated protons are converted
to more energetic, $\gamma\rightarrow(\gamma_0\gamma)^{1/2}$.
Their distribution function becomes equal (Istomin \& Sol 2009)
$f'(\gamma) = 2 const\cdot\gamma_0^ {\beta-1} \gamma^{-2\beta +1},
\, \gamma_0^{1/2} < \gamma < (\gamma_0 \gamma_1)^{1/2}$. Thus,
there is another component of jet relativistic protons, their
number is equal to
\begin{equation}
N\propto\int_0^{\infty}dz\int_{\gamma_0^{1/2}}^{(\gamma_0\gamma_1)^{1/2}}
\gamma_0^{\beta-1}\gamma^{-2\beta+1}d\gamma, \, \gamma_0=
\gamma_m^2(1+z^2/r_L^2)^{-1}.
\end{equation}
Transforming the integration area in (6), we get
\begin{eqnarray}
N\propto\int_1^{\gamma_m}\gamma^{-2\beta+1}d\gamma
\int_{(\gamma_m^2/\gamma^2-1)^{1/2}}^{(\gamma_m^2\gamma_1/\gamma^2-1)^{1/2}}
\left(1+\frac{z^2}{r_L^2}\right)^{1-\beta}\frac{dz}{r_L}+ \nonumber \\
\int_{\gamma_m}^{\gamma_m\gamma_1^{1/2}}\gamma^{-2\beta+1}d\gamma
\int_0^{(\gamma_m^2\gamma_1/\gamma^2-1)^{1/2}}
\left(1+\frac{z^2}{r_L^2}\right)^{1-\beta}\frac{dz}{r_L}.
\end{eqnarray}
The first term in Eq. (7) corresponds to relativistic protons with energies
$\gamma < \gamma_m$ similar to protons (4), accelerated from the thermal
gas. Their distribution function equals
$$
F(\gamma)=const\cdot\gamma^{-2\beta+1}\int_{(\gamma_m^2/\gamma^2-1)^{1/2}}^{(\gamma_m^2\gamma_1/\gamma^2-1)^{1/2}}
\left(1+\frac{z^2}{r_L^2}\right)^{1-\beta}\frac{dz}{r_L}.
$$
In the energy range $\gamma<<\gamma_m$ this distribution
has the same power law spectrum (4), $F(\gamma)\propto\gamma^{-2}$.
But since the number of accelerated particles in the
magnetosphere of the black hole is much less than that of thermal particles,
the contribution of these particles into the total distribution at
$\gamma < \gamma_m$ can be neglected.
The second term in Eq. (7) describes the distribution of relativistic
protons at $\gamma < \gamma_m < \gamma_m\gamma_1^{1/2}$
\begin{eqnarray}
F(\gamma)=const\cdot\gamma^{-2\beta+1}\int_0^{(\gamma_m^2\gamma_1/\gamma^2-1)^{1/2}}
\left(1+\frac{z^2}{r_L^2}\right)^{1-\beta}\frac{dz}{r_L}= \nonumber \\
\frac{1}{2}const\cdot\gamma^{-2\beta+1}\left[B(1,\beta-3/2,1/2)-B(\gamma^2/
(\gamma_m^2\gamma_1),\beta-3/2,1/2)\right].
\end{eqnarray}
Here $B (x,a,b) = \int_0^x t^{a-1}(1-t)^{b-1}dt$ is the
incomplete Beta function, $B(1,\beta-3/2,1/2) =\pi^{1/2}
\Gamma(\beta-3/2)/\Gamma (\beta-1)$, $\Gamma(x)$ is the Gamma function.
The distribution (8) with $\beta = 1.7, \, \gamma_m = 3.2\cdot 10^6$ and
$\gamma_1 = 10^5$ is shown on Figure 3. For energies $\gamma < \gamma_m
\gamma_1^{1/2}$ the distribution of relativistic protons is the power law
with index $-(2\beta-1)$. At $\gamma\simeq\gamma_m\gamma_1^{1/2}$
the distribution falls down, for $\gamma_m = 3.2\cdot 10^6$ and $\gamma_1 =
10^5$ the maximum energy is $\simeq 10^{18}$ eV.
\begin{figure}
\begin{center}
\includegraphics [width = 8 cm] {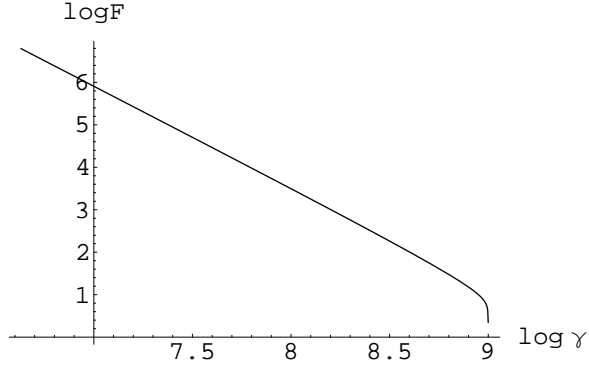}
\end{center}
\caption {The distribution function of relativistic protons in the
jet, $\gamma > \gamma_m$. The value of $\gamma_m$ corresponds to the break
in the cosmic ray spectrum. The spectral index equals -2.4.
The maximum energy is $10^{18}$ eV.}
\end{figure}

We have chosen the value of $\beta = 1.7$ from the fact that
indices of spectrum of cosmic rays before and after the break at
energy $3\cdot 10^{15}$  eV differ on the value of 0.4 -- the
spectrum becomes more soft with the index $-3.1$. The same
difference in indices must be in the source of cosmic rays, which
in our case is the relativistic jet. At $\beta = 1.7$ the index of
the spectrum (8) equals -2.4, while the index of the spectrum (4)
is equal to -2. It should be noted that the distribution function
of relativistic protons (4) at energies $\gamma < \gamma_m$, and
(8) at energies $\gamma >\gamma_m$, is continuous, i.e. $F(\gamma
= \gamma_m-0) = F(\gamma = \gamma_m+0)$. This is because the
acceleration on the light surface undergo protons of the black
hole magnetosphere with a unique spectrum -- thermal at low
energies, turning into the tail of fast particles up to
the energy $\gamma =\gamma_1$. If their energy distribution
function is $F_i({\cal E}_i)$, then the distribution of particles,
accelerated on the light surface, have also the continuous
distribution $F({\cal E}) = F_i[{\cal E}_i({\cal E})] d{\cal
E}_i/d {\cal E,} \, {\cal E}_i = {\cal E}^2/\gamma_0 m_pc^2$.

Here and hereafter, we talk about protons, bearing in mind that
they are 'heavy' particles, unlike electrons. Nuclei of $m$ mass
and $Ze$ charge will be accelerated effectively also on the light
surface. As we saw above, the efficiency of acceleration depends
on the value of $\gamma_0^{1/2} = (\omega_c/\Omega_F)^{1/2}$.
Thus, nuclei will receive energy per nucleon $(Zm_p/m)^{1/2}\simeq
2^{-1/2}$ times less than protons.

\section{Jet's remnants}
The center of the Galaxy, being active, created a relativistic
jet, particles of which spread in the Galaxy, forming isotropic
background of cosmic rays. If one consider that the angular
momentum of the massive black hole in the center of the Galaxy
coincides with that of the Galaxy, than the direction of the jet
propagation is perpendicular to the plane of the Galaxy. The jet
length can be quite large, so jet in M87 extends $\simeq 2$ kpc.
Therefore, above and below the Galactic plane (if we have two
almost symmetrical jets) one can see traces of the jet.

Let us consider a simple diffusive model of propagation of relativistic
particles, whose source is located in the center of the Galaxy $({\bf r} = 0)$,
in an interstellar medium
\begin{equation}
\frac{\partial N}{\partial t}+{\bf u}\nabla N-\nabla\hat{D}\nabla N=
Q(t)\delta({\bf r}).
\end{equation}
Here ${\bf u}$ is the velocity of the interstellar matter. We
consider the region outside the stellar galactic disk, then the
velocity ${\bf u}$ is the speed of the galactic wind, which is
along the coordinates $z$ which is perpendicular to the plane of
the Galaxy. The value of $\hat{D}$ is the diffusion coefficient,
which can be anisotropic. Diffusion depends on the intensity of
the magnetic field $B$, falling exponentially with the distance
$z$ from the galactic plane, $B = B_0\exp (-z/z_1), \, z_1 \simeq
2$ kpc. Diffusion of charged particles is determined by their
motion in the magnetic field, and the diffusion coefficient is
inversely proportional to the magnetic field strength, $D\propto
B^{-\alpha}$. So, for the most strong  Bohm diffusion $D =
cr_c/3$, $r_c$ is the proton cyclotron radius, $\alpha = 1$. Thus,
the particle diffusion increases exponentially with the coordinate
$z$, $D = D_0\exp (z/z_0), \, z_0 = z_1/\alpha$. Such a strong
dependence of the diffusion on the coordinate $z$ leads to
effective non diffusion expansion of particles along this
coordinate and decreasing its density. To take into account this
effect we  insert the function $\varphi = N\exp(z/z_0)$. Eq. (9)
becomes
\begin{eqnarray}
\exp\left(-\frac{z}{z_0}\right)\frac{\partial\varphi}{\partial t}+
u\exp\left(-\frac{z}{z_0}\right)\left(
\frac{\partial\varphi}{\partial z}-\frac{\varphi}{z_0}\right)+\frac{D_0}{z_0}
\frac{\partial\varphi}{\partial z}- \nonumber \\
D_0\left[\frac{\kappa}{\rho}\frac{\partial}
{\partial\rho}\left(\rho\frac{\partial\varphi}{\partial\rho}\right)+
\frac{\partial^2\varphi}{\partial z^2}\right]=Q(t)\frac{\delta(\rho)\delta(z)}
{2\pi\rho}.
\end{eqnarray}

We consider the distribution of particles as azimuthal symmetric
depending on the distance $z$ and the cylindrical radius $\rho$.
The value of $\kappa$ is the ratio of the transverse diffusion
coefficient $D_\perp$, perpendicular to $z$, to the longitudinal
one $D_\parallel$, along $z$, $\kappa=D_\perp/D_\parallel$. We see
that in the equation of particle motion there appears the
effective velocity along $z$, $u_0 = D_0/z_0$. It arises as a
result of the exponential growth of the particle diffusion over $z$.
For characteristic values of $D_0=5\cdot 10^{28}$ $cm^2$/s and
$z_0=2$ kpc the velocity $u_0$ is of $u_0\simeq 10^2$ km/s. This
speed is much larger than the speed of the galactic wind $u$,
which at distances of several kpc from the galactic plane is less
than $30$ km/s (Ptuskin, 2007). Although the wind velocity
increases with distance $z$, the exponential factor $\exp(-z/z_0)$
in Eq. (10), allows us to ignore the velocity of the galactic wind
in comparison with the velocity $u_0$. Values of $D_0$ and $z_0$
specify scales of length and time, so it is convenient to go to
the dimensionless variables in Eq. (10) $z'=z/z_0, \,
\rho'=\rho/z_0, \, t'=t(D_0/z_0^2), \, \varphi'=\varphi z_0^3$.
Eq. (10) becomes (primes are omitted)
\begin{equation}
\exp(-z)\frac{\partial\varphi}{\partial t}+
\frac{\partial\varphi}{\partial z}-\frac{\partial^2\varphi}{\partial z^2}
-\frac{\kappa}{\rho}\frac{\partial}
{\partial\rho}\left(\rho\frac{\partial\varphi}{\partial\rho}\right)
=Q(t)\frac{\delta(\rho)\delta(z)}{2\pi\rho}.
\end{equation}
At $z >1$ the propagation (first derivative over $z$)
predominates over the diffusion (second derivative), and Eq. (11)
makes easy
$$
\varphi=\frac{Q[t-(1-e^{-z})]}{4\pi\kappa z}\exp\left(-\frac{\rho^2}{4\kappa z}
\right).
$$
From this solution one can see that to the point $z$ come particles, which were
emitted by the source at the retarded time $t' = t-(1-e^z)$. If the
time of the jet's start is $t = 0$, particles will lift to the maximum height
of $z_m =-\ln(1-t),\, t < 1$.  Formally at $t = 1$
particles come to infinity during the finite time, that is impossible.
The velocity limitation implies the condition $z_m<(cz_0/D_0)t$, $c$ is the
speed of the light, which is not difficult to hold because
$cz_0/D_0\simeq 3.6\cdot 10^3$. Knowing the value of $z_m$
from observation one can estimate the time $t_1$ when the power source of cosmic
rays in the center of the Galaxy begins to work, i.e. when the jet starts,
$t_1 = 1-\exp(z_m)$. In dimensional units $t_1=t_0[1-\exp(z_m/z_0)], \,
t_0 = z_0^2/D_0$.
When $z_0 = 2$ kpc and $D_0 = 5\cdot 10^{28}\, cm^2/s$ the time $t_0$ is
$t_0 = 7.6\cdot 10^{14} \, s = 2.4\cdot 10^7$ yr. The gamma radiation observed
above and below the galactic plane extends to the height of
about 8 kpc (Su et al., 2010), i.e. $z_m\simeq  4$. This means that in
fact $t_1=t_0$.
In addition, one can find the time $t_2$ when the source
turns off. If the jet worked a short time, all particles would
have lifted in height at $z\geq 1$, and we would see their absence
near the galactic plane. Since this is not observed in bubbles, then
$t_2 < t_0(1-1/e)\simeq 0.6 t_0 $.

Knowing the solution for $\varphi$ at $z>1$ we can find the density
of relativistic particles $N (t,z,\rho)$ in the same region
\begin{equation}
N=\frac{Q[t-(1-e^{-z})]}{4\pi\kappa z_0^3}\exp\left[-\left(\frac{\rho^2}
{4\kappa z}+z+\ln(z)\right)\right].
\end{equation}
The distribution $N(z,\rho)$ (12) for the
permanent source, $Q = const(t)$, is shown on Figure 4.
\begin{figure}
\begin{center}
\includegraphics [width = 8 cm] {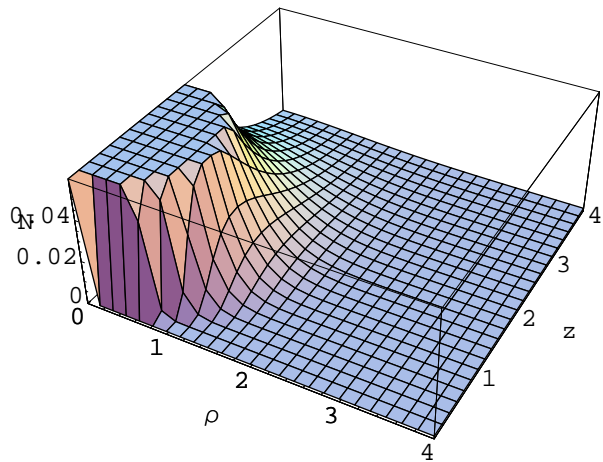}
\end{center}
\caption{The density distribution of relativistic particles (12)
above the galactic plane in dimensionless coordinates: $z$, the distance
from the plane of the Galaxy, and $\rho$, the cylindrical radius.}
\end{figure}
We also draw levels of the constant density $N, \, \rho^2/4\kappa z
+z+ln(z)=const$, Figure 5.
\begin{figure}
\begin{center}
\includegraphics [width = 6cm] {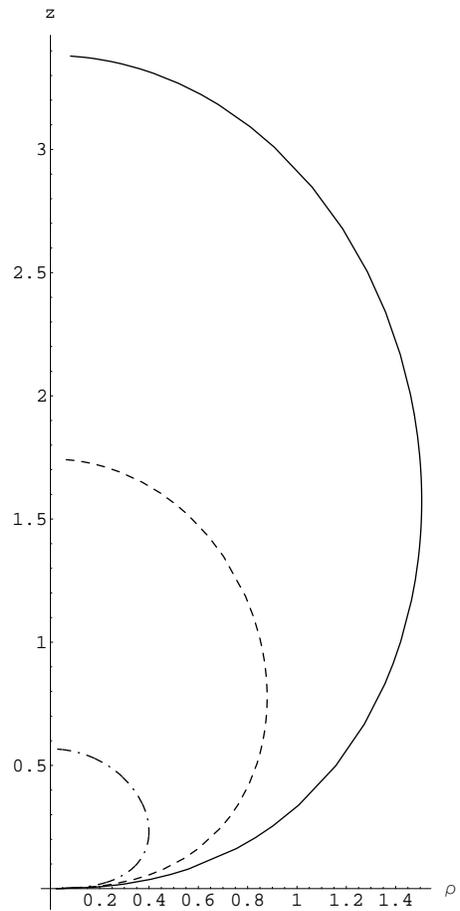}
\end{center}
\caption {Levels of the constant density of relativistic
particles specified by the distribution (12).}
\end{figure}
The value of $ \kappa $, the anisotropy of diffusion, is chosen to
be equal to $\kappa = 0.14$ from those considerations that the
observed ratio of bubble's scales  $z/\rho\simeq 8 kpc/3 kpc =
8/3$ would be consistent with the forms of the constant density
profiles painted on Figure 5. It seems that such value of $\kappa$
indicates that the magnetic field in the halo of the Galaxy near
the center is mostly vertical. And this is natural because for the
cylindrical symmetry radial and azimuthal magnetic field
components should approach zero on the axis $\rho = 0$.

\section{Discussion}

We have shown that whereas in the past the center of the Galaxy was
active and radiated the jet, its energy and composition are sufficient to
explain the origin of cosmic rays in the Galaxy. Bubbles of a
relativistic gas observed in gamma, x-ray and radio bands above and below
the galactic plane, apparently, are remnants of the bipolar jet emitted from
the vicinity of the massive black hole in the center of the Galaxy.
The vertical size of the bubble $z\simeq 8$ kpc permits us to
estimate the time of switching on of the jet, $t_1\simeq t_0 = 2.4\cdot 10^7$
years ago. We can also estimate the lower limit of the jet's work,
$\Delta t = t_1-t_2 > 0.4 t_0\simeq 10^7$ yr. During the time $\Delta t$ the
jet got the energy $L_J\Delta t\simeq 8.9\cdot 10^{45}$ erg/s
$\times$ $3\cdot 10^{14}$ s $\simeq 2.7\cdot10^{60}$ erg,
that is slightly larger than the energy stored in the black hole rotation
$\simeq 10^{60}$ erg. However, if we estimate the mass of the accreted
matter, absorbed by the black hole at the same time,
it can reach a large part of the mass of the black hole,
$\Delta M\simeq {\dot M}_{Edd}\Delta t=
9.2\cdot 10^{-2}M_{\odot}\,yr^{-1}\times 10^7\,yr
\simeq 10^6 M_{\odot}$, $\Delta M\simeq M/4$.
Transmitted to the black hole by the accreted matter, the angular
momentum $ \Delta J$ can even exceed its initial value (Istomin, 2004).
The jet's energy $\simeq 10^{60}$ erg is enough to fill by cosmic rays
as the disk ($10^{55}$ erg), as the halo ($10^{57}-10^{58}$ erg) of the Galaxy.
Filling of the disk of the Galaxy by relativistic particles is described
by the same Eq. (11), but there should be $|z| < 1$. Therefore, we can ignore
the dependence of the diffusion coefficient on the distances $z$, and
solve the pure diffusion equation
$$
\frac{\partial N}{\partial t}
-D_\parallel\frac{\partial^2 N}{\partial z^2}
-D_\perp\frac{1}{\rho}\frac{\partial}
{\partial\rho}\left(\rho\frac{\partial N}{\partial\rho}\right)
=Q(t)\frac{\delta(\rho)\delta(z)}{2\pi\rho}.
$$
Here we are interested in distribution of particles in the disk on
the transverse distances $\rho$, so we average this equation
over $z$ and get
\begin{equation}
\frac{\partial {\bar N}}{\partial t}-\frac{D_g}{\rho}\frac{\partial}
{\partial\rho}
\left(\rho\frac{\partial {\bar N}}{\partial\rho}\right)=Q(t)\frac{\delta(\rho)}
{2\pi\rho},
\end{equation}
where the value of $D_\perp = D_g$ is the diffusion coefficient of cosmic rays
in the galactic disk. The solution is
$$
{\bar N}=\frac{1}{4\pi D_g}\int_0^{t_1}\frac{1}{\tau}\exp\left(-\frac{\rho^2}
{4D_g\tau}\right)Q(t_1-\tau)d\tau.
$$
The time $t_1$ is the start time of the jet. If the jet had worked with
constant power $Q$ and switched off at the time $t_2$, then the density
distribution of cosmic rays in the disk is
\begin{equation}
{\bar N}=\frac{Q}{4\pi D_g}\int_{\rho^2/4D_gt_1}^{\rho^2/4D_g(t_1-t_2)}
x^{-1}e^{-x}dx.
\end{equation}
The solution (14) shows that if $\rho^2/4D_gt_1< 1$ then the distribution of
the density over the radius $\rho$ is almost uniform. For
$\rho^2/4D_g(t_1-t_2) > 1$ it is logarithmic, ${\bar N}\propto -\ln(\rho^2/
4D_gt_1)$, and for $\rho^2/4D_g(t_1-t_2) < 1$ it is constant,
$N\propto \ln(t_1/(t_1-t_2))$. The diffusion coefficient
of cosmic rays in the disk equals
$D_g = 2.2\cdot 10^{28}\gamma^{0.6}\,cm^2/s$ (Ptuskin, 2007).
The condition $R^2/4D_g < t_0, \, R\simeq 15$ kpc is the disk radius,
imposes a lower limit on the energy of protons, homogeneously filling
the galactic disc, $\gamma > 400$. Apart the diffusion
particles can move freely along the regular magnetic field of spiral
arms of the galactic disk. The necessary for filling velocity $R/t_0 \simeq
6\cdot 10^7$ cm/s $=2\cdot 10^{ -3}$ c does not contradict
the observed anisotropy of cosmic rays $\delta$, $\delta\simeq 10 ^{-3}$.
Since the dependence of the distribution (14) over the energy is
determined not only by the
energy spectrum of the source $Q(\gamma)$ but also the dependence of the
diffusion coefficient $D_g\propto\gamma^{0.6}$ over the energy, the spectrum
of particles in the disk will be softer than that in
the source. Thus, the discussed mechanism of origin of galactic cosmic rays
by the jet, emitted from the center of the Galaxy, satisfactorily explains the
observed spectrum, the index -2.7 (-2.6 for the jet) before the break,
and the index -3.1 (-3.0 for the jet) after the break.

Cosmic rays, filling simultaneously the galactic disk and the
halo, flow out from the Galaxy. Their lifetime $\tau$ is
determined by as the energy loss time $\tau_E$, as the time of
diffusion leakage from the Galaxy after the source of relativistic
particles stopped, $\tau_D$. The time $\tau_E$ is estimated as
$\tau_E\simeq 3\cdot 10^7$ yr (Strong \& Moskalenko, 1998). It is
larger than the time of the jet's start $t_1$, i.e. beginning of the
filling of the Galaxy by cosmic rays, $\tau_E>t_1$. The diffusion
time $\tau_D = r^2/4D$  is determined by the distance $r$ that
particles travel during the period from the switching on of the
source $r = (4Dt_1)^{1/2}$. Thus, $\tau_D\simeq t_1$  does not
depend on the energy of particles. This time is also larger than
the time of jet's switching off $t_2$, $t_2< 0.6t_1$. We see that to now
the distribution of relativistic particles, generated by the jet,
does not change noticeable.

\section*{Aknowlegements}
This work was done under support of the Russian Foundation for Fundamental
Research (grant number 11-02-01021).

\end{document}